\documentclass[useAMS,usenatbib]{mn2e}

\usepackage{graphicx}
\usepackage{epsfig}

\begin{document}
\title{The Extreme Ultraviolet Spectrum of the Kinetically Dominated Quasar 3C 270.1}
\author[Brian Punsly and Paola Marziani] {Brian Punsly and Paola Marziani\\ 1415 Granvia Altamira, Palos Verdes Estates CA, USA
90274 and ICRANet, Piazza della Repubblica 10 Pescara 65100, Italy\\
\\INAF, Osservatorio Astronomico di Padova, Italia.\\
 \\E-mail: brian.punsly1@verizon.net}

\maketitle \label{firstpage}
\begin{abstract}  Only a handful of quasars have been identified as
kinetically dominated, their long term time averaged jet power,
$\overline{Q}$, exceeds the bolometric thermal emission, $L_{bol}$,
associated with the accretion flow. This letter presents the first
extreme ultraviolet (EUV) spectrum of a kinetically dominated
quasar, 3C 270.1. The EUV continuum flux density of 3C 270.1 is very
steep, $F_{\nu} \sim \nu^{-\alpha_{EUV}}$, $\alpha_{EUV} =2.98\pm
0.15$. This value is consistent with the correlation of
$\overline{Q}/L_{bol}$ and $\alpha_{EUV}$ found in previous studies
of the EUV continuum of quasars, the EUV deficit of radio loud
quasars. Curiously, although ultraviolet broad absorption line (BAL)
troughs in quasar spectra are anti-correlated with $\overline{Q}$,
3C 270.1 has been considered a BAL quasar based on an SDSS spectrum.
This claim is examined in terms of the EUV spectrum of OVI 1and the
highest resolution CIV spectrum in the archival data and the SDSS
spectrum. First, from [OIII]4959,5007 (IR) observations and the UV
spectral lines, it is concluded that the correct redshift for 3C
270.1 is 1.5266. It is then found that the standard measure of broad
absorption, BALnicity = 0, for MgII 2800, CIV 1549 and OVI 1032 in
all epochs.
\end{abstract}
The quasar 3C 270.1 contains one of the most powerful jets of any
known active galactic nucleus. The jet power greatly exceeds any
other associated with a quasar that has been claimed to have an
ultraviolet (UV) broad absorption line. These extreme properties of
3C 270.1 offer a unique laboratory for exploring the mechanism of
relativistic jet formation. The long term time averaged jet power,
$\overline{Q}$, determined from low frequency radio lobe emission,
is so extreme that the quasar was classified as kinetically
dominated $\overline{Q}/L_{bol}>1$, where $L_{bol}$ is the
bolometric thermal emission associated with the accretion flow
\citep{pun07}. The physical relevance of this result is uncertain
since the plasma in the radio lobes was ejected from the central
engine $>10^{5}$ years before the UV emitting gas reached the
environs of the central black hole. This letter uses the EUV
spectrum of 3C 270.1 to address two fundamental issues. First, can a
quasar actually emit more power in the jet than is being radiated as
thermal emission? Secondly, why are broad absorption line winds
anti-correlated with jet emission on super-galactic scales? Both of
the issues provide valuable clues to the jet launching mechanism.
\par Even though the radio lobes of powerful radio loud quasars (RLQs) are located light travel times
typically $10^{5}$ - $10^{6}$  yrs from the central black hole, they
tend to be connected by a radio jet. Quasars with radio lobes on
super-galactic scales are very rare, $\sim 1.7\%$ of all quasars
have such extended structure \citep{dev06}. The existence of the
connective bridge formed by the jet implies that the energy source
seems persistent for long periods of time. It is a mystery how the
jet can be powered for so long and it is unclear how much the jet
power fluctuates over $10^{6}$ yrs compared to its average value. In
\cite{pun14,pun15}, it was shown that $\overline{Q}/L_{bol}$ (which
depends on a long term average) was correlated with the deficit of
EUV emission quantified by $\alpha_{EUV}$ (the flux density scales
as $F_{\nu}\propto\nu^{-\alpha_{EUV}}$ ). The EUV is believed to be
the putative Wien tail of the optically thick emission from the
innermost region of the accretion flow; a region of the flow that
interacts in real time with the central black hole \citep{pun15}.
Like the bridge formed by the jet, this also seems to indicate a
rather persistent dynamic near the central black hole that is
sustained for most of $\sim10^{6}$ yrs of the radio source lifetime.
Furthermore, there is only a modest degree of scatter from the
average trend in the $\overline{Q}/L_{bol}$ - $\alpha_{EUV}$ plane
(see Figure 2) for the RLQ population. The scatter plot samples
quasars at random epochs in their radio loud lifetime. This modest
scatter indicates that the fluctuations in the dynamics are
typically not that large. Some ``baseline" dynamical configuration
seems to exist in each quasar more often than not for the majority
of these objects. In the next section, this is interpreted with
other evidence to indicate that 3C 270.1 had a powerful jet,
$Q(t)/L_{bol}\sim 1$, when the EUV radiation was emitted.
\par Another aspect of this study is to analyze the claim of
\cite{gib09} that 3C 270.1 has broad absorption in CIV and not
associated absorption. This conclusion contradicts previous findings
\citep{and87}. This is physically significant since there are only a
handful of known RLQs with extended structure on super-galactic
scales that are bona-fide BALQSOs as defined by the BALnicty index.
None of these have $\overline{Q}/L_{bol}$ within a factor of 10 of
3C 270.1. This conflict is analyzed in Section 2. Quantifying the
jet power and the interplay between the BAL wind suppression and jet
power provides fundamental insight into the nature of the jet
launching in quasars.
\begin{figure}
\includegraphics[width=79 mm, angle= 0]{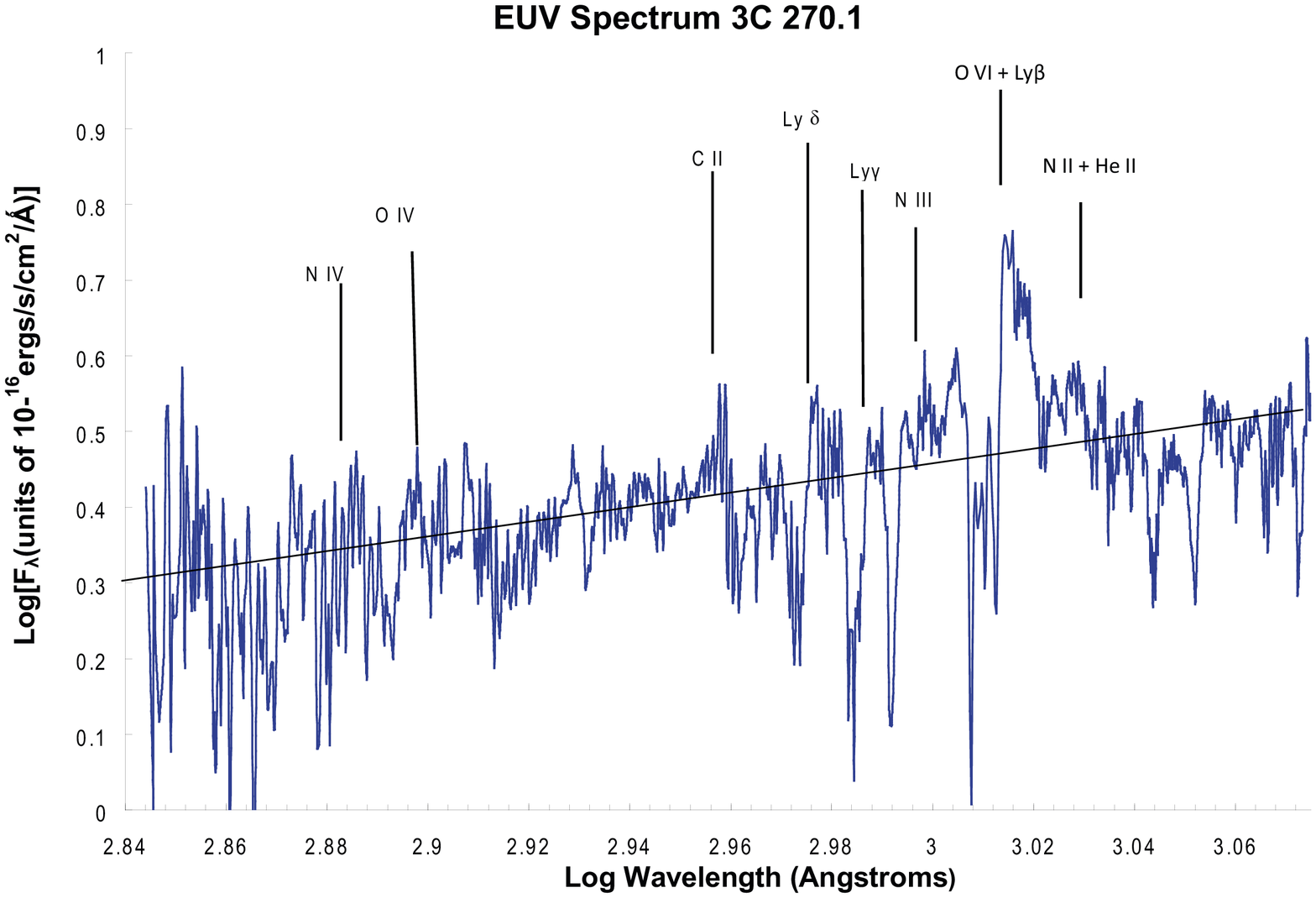}
\includegraphics[width=79 mm, angle= 0]{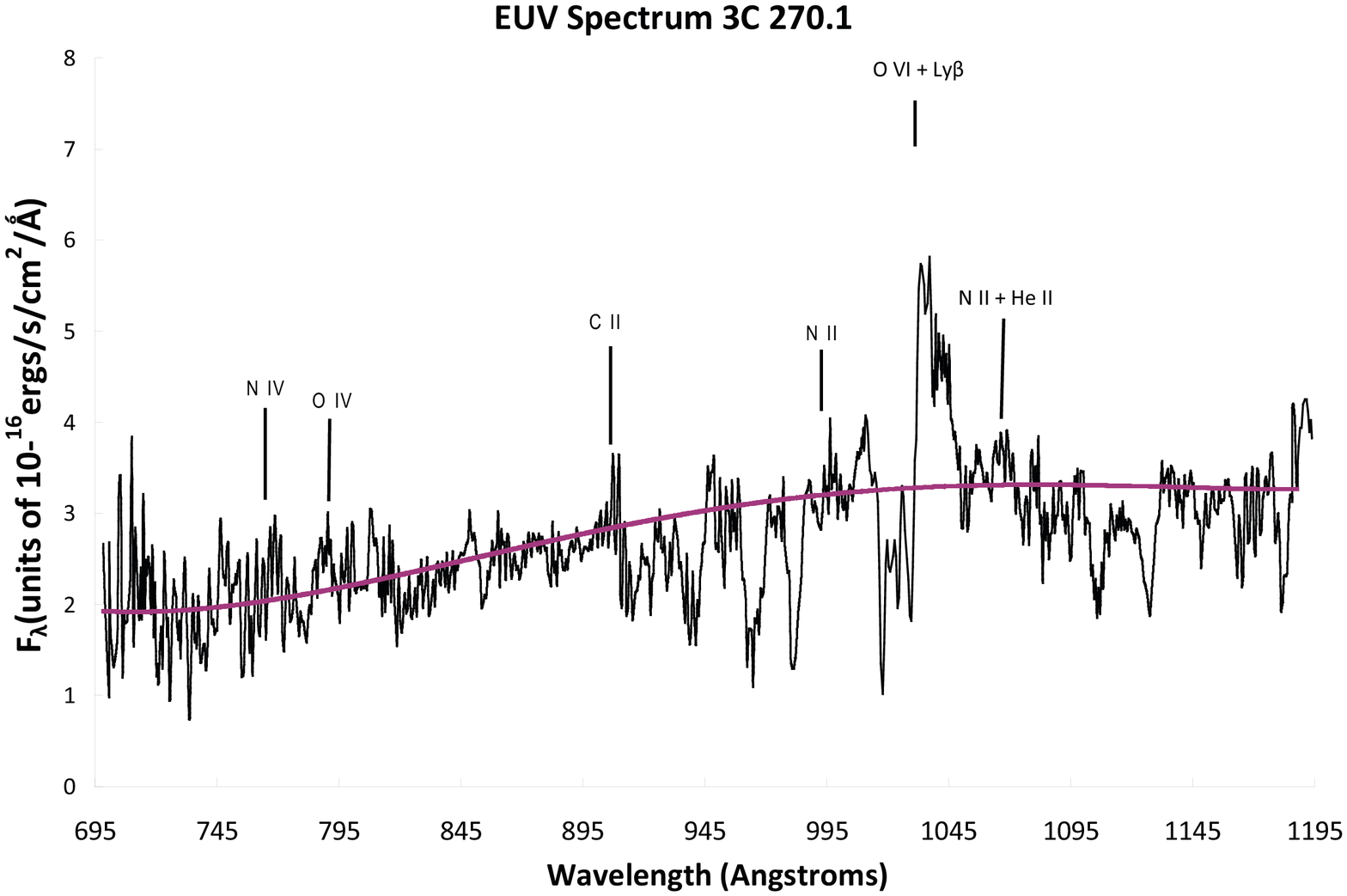}
\caption{The EUV spectrum of 3C 270.1. The top frame is the power
law fit (solid black line) to the continuum. The bottom frame is a
fourth order polynomial fit (solid red line) to the continuum.}
\end{figure}

\section{The EUV Spectrum and Jet Power} 3C 270.1 was observed on 7/24/2000 for
2380 s with the Hubble Space Telescope (HST) with the STIS
spectrograph and G230L grating. The data was downloaded form MAST
and corrected for Galactic extinction with the CCM absorption law
\citep{car89}. The absorption from the Ly$\alpha$ valley was
de-convolved from the data using the empirical statistical methods
of \citep{zhe97}. An intervening Lyman limit system near the quasar
systemic redashift was removed with the decrement method
\citep{shu12,pun14}. Namely, the Lyman limit system was removed by
assuming a single cloud with a $\nu^{-3}$ opacity and an absorption
that was scaled by the decrement between the continuum and the
trough depth. This was considered acceptable because the power law
above the Lyman limit system continues smoothly through the
corrected region (see Figure 1). The power law between $1100 \AA$
and $700 \AA$ is defined by $\alpha_{EUV} = 2.98 \pm 0.15$. The top
frame of Figure 1 is the power law fit (solid black line) to the
continuum. The bottom frame is a fourth order polynomial fit (solid
red line) that provides an upper limit to the continuum in the
analysis of OVI absorption in the next section.
\par A method that allows one to convert 151 MHz flux densities,
$F_{151}$ (measured in Jy), into estimates of long term time
averaged jet power, $\overline{Q}$, (measured in ergs/s) is given by
Equation (1) \citep{wil99,pun05}:
\begin{eqnarray}
 && \overline{Q} \approx [(\mathrm{\textbf{f}}/15)^{3/2}]1.1\times
10^{45}\left[X^{1+\alpha}Z^{2}F_{151}\right]^{0.857}\mathrm{ergs/s}\;,\\
&& Z \equiv 3.31-(3.65)\times\nonumber \\
&&\left[X^{4}-0.203X^{3}+0.749X^{2} +0.444X+0.205\right]^{-0.125}\;,
\end{eqnarray}
where $X\equiv 1+z$, $F_{151}$ is the total optically thin flux
density from the lobes. Deviations from the overly simplified
minimum energy estimates are combined into a multiplicative factor,
\textbf{f}, that represents the small departures from minimum
energy, geometric effects, filling factors, protonic contributions
and low frequency cutoff \citep{wil99}. In \cite{blu00}, it was
argued that $10 < \mathrm{\textbf{f}}< 20$. Alternatively, there is
another isotropic estimator in which the lobe energy is primarily
inertial in form \citep{pun05}:
\begin{eqnarray}
&&\overline{Q}\approx
5.7\times10^{44}(1+z)^{1+\alpha}Z^{2}F_{151}\,\mathrm{ergs/sec}\;.
\end{eqnarray}
Define the radio spectral index, $\alpha$, as
$F_{\nu}\propto\nu^{-\alpha}$. Equation (1) with
$\mathrm{\textbf{f}} = 20$ is the maximum upper bound on
$\overline{Q}$ and Equation (2) is the lower bound $\overline{Q}$
that is used in the following. In this paper, we adopt the following
cosmological parameters: $H_{0}$=70 km/s/Mpc, $\Omega_{\Lambda}=0.7$
and $\Omega_{m}=0.3$. The radio images in \cite{gar91} indicate that
this is a lobe dominated quasar. Therefore, one expects the 151 MHZ
flux density (14.96 Jy from the NASA Extragalatic Database) to be
dominated by the radio lobe emission. Thus, from Equations (1) - (3)
that $\overline{Q} = 7.5 \times 10^{46} \pm 1.5 \times 10^{46}$
erg/s.
\par Since the data used here covers the
peak of the SED at $\lambda \approx 1100 \AA$, an accurate
expression, $L_{\mathrm{bol}} \approx 3.8 \lambda L_{\lambda}
(\lambda =1100 \AA)$, can be used to estimate the accretion disk
luminosity. This does not include reprocessed IR emission in distant
molecular clouds \citep{dav11,pun14}. From the flux density in
Figure 1,
 $L_{bol} = 5.2 \times 10^{46} \pm 0.5 \times 10^{46}$ erg/s.
 Combining this with the estimate of $\overline{Q}$, above, $\overline{Q}/L_{bol}=1.4$.
 3C 270.1 is kinetically dominated in this context.

\par The top frame of Figure 2 places 3C 270.1 on the same scatter plot of
$\log[\overline{Q}/L_{bol}]$ and $\alpha_{EUV}$ as the other quasars
for which a similar data reduction was performed in Figure 3 of
\cite{pun15}. 3C 270.1 has the most powerful jet and it lies at the
end of the extrapolated trend of the other powerful RLQs. Figure 2
relates real time properties (meaning concurrent with the epoch that
the data was sampled in the quasar rest frame), $L_{bol}$ and
$\alpha_{EUV}$, with a long term time average property,
$\overline{Q}$. The coefficient of determination to a linear fit is
0.5927 so the scatter is modest. Each data point in Figure 2 is a
random snapshot in time of the properties of the innermost accretion
flow during the lifetime each RLQ. The degree of scatter indicates
that there is a baseline or fiducial configuration of the central
engine in each RLQ and the fluctuations over time from this
configuration are usually ``modest" and large fluctuations are not
typical. This concept is quantified below in order to see if there
is predictive power of real time jet parameters from the data in
Figure 2. To begin with, segregate the real time variables from
$\overline{Q}$ by inverting the linear fit to the data in the top
frame of Figure 2. This process yields a real time variable
\begin{equation}
G(t) = 2.58\alpha_{EUV}(t)+ \log{[3.8 \lambda L_{\lambda} (\lambda
=1100 \AA)(t)/10^{45} \mathrm{ergs/s}]}\;.
\end{equation}
As an alternative to a scatter plot in the
$\log[\overline{Q}/L_{bol}]$ - $\alpha_{EUV}$ plane (the top frame
of Figure 2), one can also consider the data scatter in the
$\log[\overline{Q}]$ - $G(t)$ plane. Define $\overline{\Psi}$ as the
best fit power law estimator of $\log(\overline{Q})$ in the new
scatter plane. Then
\begin{equation}
\overline{\Psi} = 38.97 G(t)^{0.0815} \;,
\end{equation}
and the coefficient of determination is 0.5937.
\par In order to extract the predictive capability of Equation (5) for the real time jet
power (that which is concurrent with the EUV measurements), $Q(t)$,
consider the following. The EUV emission and the jet launching both
originate in the immediate vicinity of the central supermassive
black hole \cite{pun15}. The fundamental assumption of this analysis
is: \emph{Since the EUV and the jet emanate from a common compact
region, the correlation found in Figure 2 between these powerful
dynamic elements is not spurious or coincidental, but results from
an actual physical connection between, $\alpha_{EUV}(t)$, $L_{bol}
(t)$ and $Q(t)$ in real time.} In particular, the average trend in
Figure 2 is a direct consequence of this real time interaction, or
alternatively stated Equation (5) is a consequence of a more
fundamental quasi-simultaneous relation
\begin{equation}
\log{Q(t)}=38.97 G(t)^{0.0815}\;.
\end{equation}
However, this relationship can only true in an average sense due to
variations from RLQ to RLQ, and epoch to epoch variations in both
the geometry as well as the physical state of the innermost
accretion flow. Describe the stochastic behavior of $\log(Q(t))$ by
a probability distribution, $\Psi\{Q[G(t)]\}$, that represents the
physics of a complicated dynamical system created by an ensemble of
numerous microphysical domains. From Equations (5) and (6), the mean
of the distribution is, $\mu[\Psi\{Q[G(t)]\}]\equiv \overline{\Psi}
= 38.97 G^{0.0815}$, where $G$ is the time average of $G(t)$. To
estimate the variance of $\Psi\{Q[G(t)]\}$, note that each data
point in Figure 2 is a random time snapshot of the inner most
accretion flow, so the best fit to the trend represents the time
averaged configuration for a given $\log(\overline{Q})$ and the
dispersion from this trend results from $G(t)$ (and therefore $Q(t)$
by Equation (6)) varying from the mean value. From the scatter in
the $\log[\overline{Q}]$ - $\log{Q(t)}$ plane, one can compute the
standard deviation, $\sigma = 0.46$. Assuming a normal distribution,
$Z$,
\begin{equation}
\Psi\{Q[G(t)]\} = Z[\mu = 38.97 G^{0.0815}, \, \sigma =0.46] \;.
\end{equation}

Based on the confidence contours (computed from $\Psi\{Q[G(t)]\}$),
plotted as dashed blue curves in the bottom frame of Figure 2, $7.3
\times 10^{45}\mathrm{ergs/s}< Q(t)< 4.8 \times 10^{47}
\mathrm{ergs/s}$ with 90\% confidence. A large $Q(t)$ is consistent
with the existence of a powerful radio core $\approx 4.9 \times
10^{44} \mathrm{ergs/s}$ \citep{lon93,aku94,gar91}.
\begin{figure}
\includegraphics[width=69 mm, angle= 0]{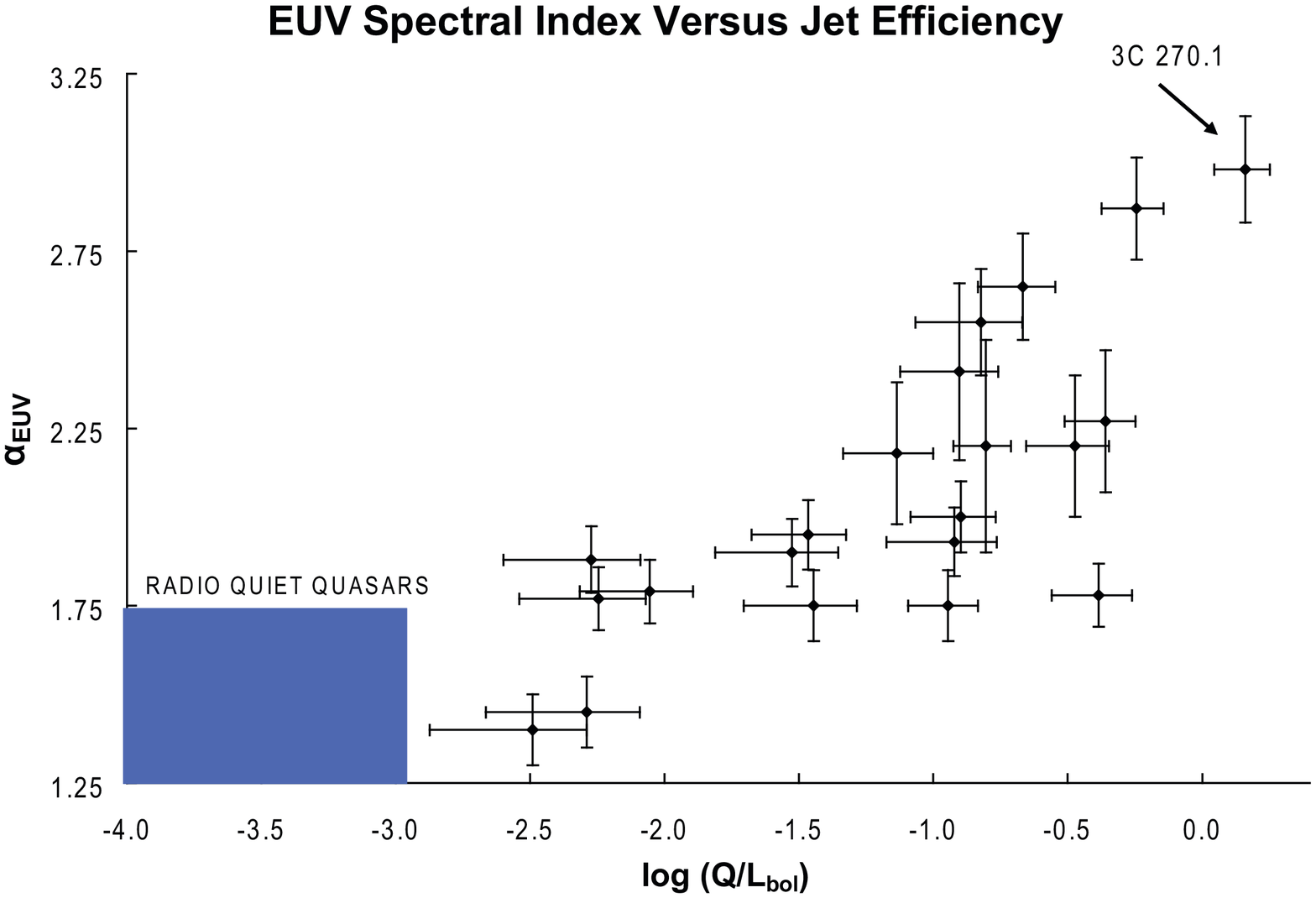}
\includegraphics[width=69 mm, angle= 0]{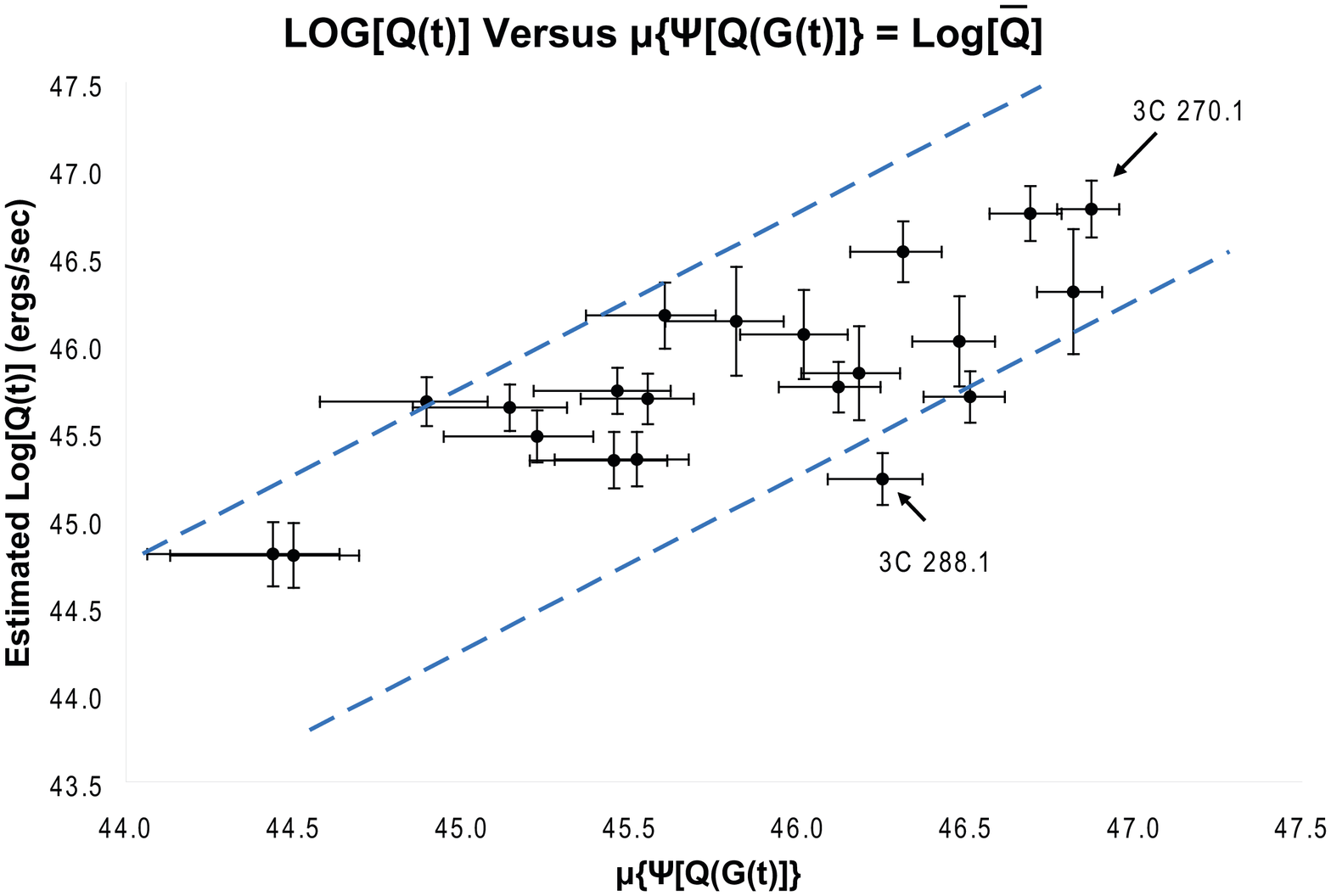}
\caption{The top frame is a scatter plot of $\overline{Q}/L_{bol}$
vs. $\alpha_{EUV}$. The bottom frame is a scatter plot of $Q(t)$
estimated from Equation (6) relative to the 90\% confidence contours
(blue dashed) of the probability distribution, $\Psi\{Q[G(t)]\}$, in
Equation (7).}
\end{figure}
\par As a verification of the validity of this
method consider, the outlier in Figure 2, 3C 288.1. The EUV deficit
is much less than expected for the value of $\overline{Q}$. However,
the radio core is rather weak with a 1 GHz - 100 GHz luminosity of
$\approx 1.0 \times 10^{43} \mathrm{ergs/s}$ \citep{rei95}. The
source has very little intrinsic absorption as evidenced by the flat
EUV spectral index, so one expects a direct line of sight to the
core. This implies that it is an intrinsically weak radio core with
no significant free-free absorption or Doppler de-boosting. The
conclusion is that 3C 288.1 is an outlier in Figure 2 because the
quasar is currently in a state of relatively weak radio activity.
\section{UV Emission Lines and Absorption Lines}
One of the most striking tends in the quasar phenomenon is that the
broad absorption line prominence is anti-correlated with the radio
loudness of the quasar \citep{bec01,sha08}. If one considers RLQs
with extended emission on super-galactic scales (radio lobes) this
anti-correlation is stronger. Only a handful of BALQSO candidates
with extended radio emission have ever been found even in deep
surveys \citep{gre06}. Considering the large jet power in 3C 270.1,
the claim of \cite{gib09} that there was a CIV BAL in the SDSS
spectrum (under the formal definition of \cite{wey91} per Equation
(8), below) is startling. Even more so because the absorption of CIV
in 3C 270.1 was analyzed by R. Weymann in \cite{and87} and was an
example of what is not a BAL, but what they call associated
absorption. Since the anti-correlation between the power of extended
radio lobes and BALs is likely related to the phenomenon of jet
launching and BAL wind formation, this controversial finding is
examined in detail in this section. The top frame of Figure 3 is an
overlay of the MMT (Multiple Mirror telescope) data from
\cite{and87} and the SDSS data (see also Table 1). The OVI
absorption is displayed in the bottom frame of Figure 3. The upper
limit of the continuum is solid red (the fourth order polynomial
fit) and the blue dashed line is the power law fit from Figure 1. It
is clearly broader and higher velocity than the CIV absorber. The
data was plotted with a smoothing window of $1.25 \AA$ in the quasar
rest frame.
\par The BALnicity index, BI, was defined in \cite{wey91} as
\begin{equation}
BI = \int_{v=-25000}^{v=-3000}  [1 -F(v)/0.9]C\, dv\;,
\end{equation}
where $F(v)$ is the flux density normalized to the continuum level
as a function of $v$, the velocity from the QSO rest frame line
emission frequency in km/s. The step function, $C(v) \neq 0$ if and
only if there more than 2000 km/s of continuous absorption beyond
-3000 km/s. This measure of broad absorption has been borne out over
the years as very robust. More lax measures of absorption (such as
associated absorption and mini-BALs) have turned out to represent
different classes of objects, that have less X-ray absorption than
BALQSOs and larger radio luminosity than bona-fide BALQSOs
\citep{sha08,pun06,kni08}. In deriving this measure, the authors
considered narrower associated absorption and decided that this was
a different phenomenon. The function $C(v)$ was designed to
segregate out this type of absorption.

\par In order to implement Equation (8), we need an accurate determination of the redshift. From the [OIII] narrow
lines observed in \citet{jac97}, we obtain $z= 1.5226$. This agrees
with our UV line analysis of the SDSS data, 1.5228 $\pm$ 0.0013\ at
the 1$\sigma$ confidence level. $C \neq 0$ requires absorption
beyond -5000 km/s.  Thus, from Figure 3 and Table 1, there is no
absorption beyond -5000 km/s and C = 0 for CIV and OVI for all $v$.
These absorption have $BI = 0$ and are examples of associated
absorption that is common in RLQs.

\begin{figure}
\includegraphics[width=54 mm, angle= 0]{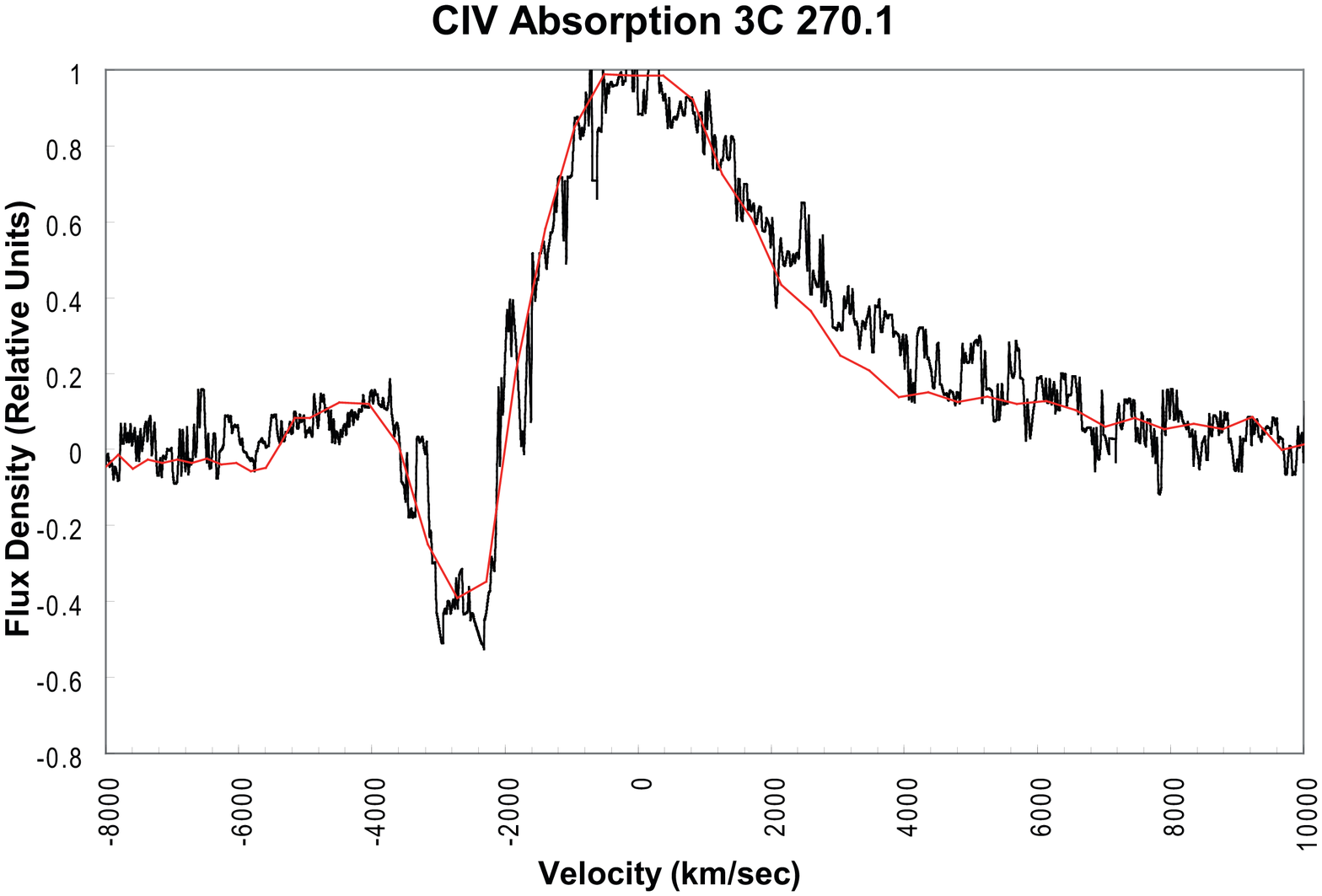}
\includegraphics[width=54 mm, angle= 0]{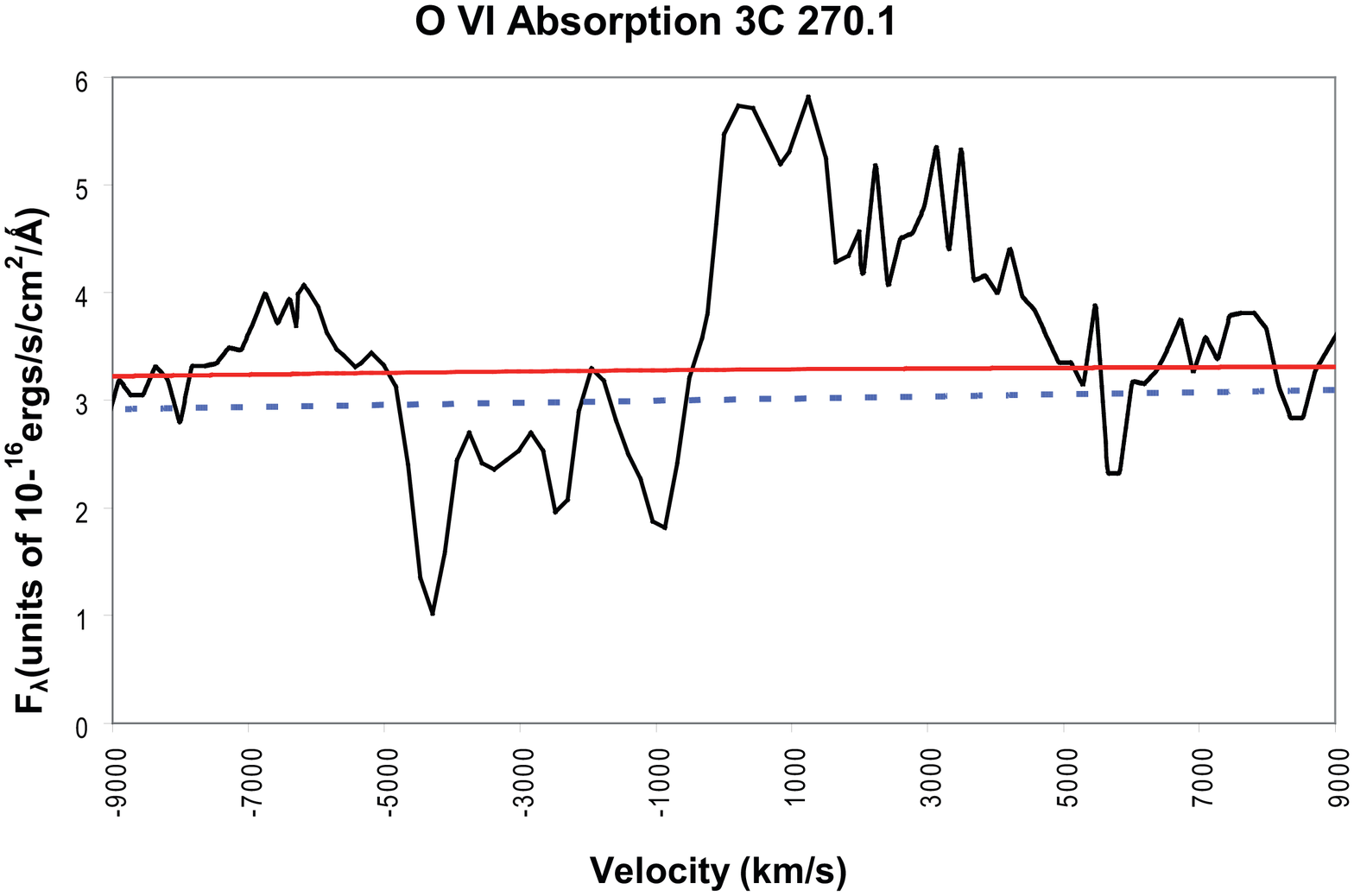}
\caption{Ultraviolet absorption lines occur in the spectrum. The CIV
line as measured by MMT is plotted in black in the top frame. The
SDSS low resolution CIV spectrum is overlayed in red. The data is
referenced to the estimated continuum level. The OVI absorption and
emission line is plotted in the bottom frame. The solid red line is
the fourth order polynomial fit and the dashed blue line is the
power law fit to the continuum. }
\end{figure}

\par OVI shows absorption maxima at
$v \approx -4290\;, -2480$ km/s and CIV shows absorption at $v
\approx -2660$ km/s. From the data in \cite{ald94} and the SDSS data
in Table 1, the narrow MgII absorption arises from gas that has an
outflow velocity of $v \approx -2620$ km/sec. This indicates the
following wind dynamic. There is a a narrow absorption line wind
with $ v_{\mathrm{wind}} \approx 2600$ km/s and second high
ionization wind with $ v_{\mathrm{wind}} \approx 4300$ km/s. It is
not clear if the high ionization wind is the base of a common wind
that coasts at $ v_{\mathrm{wind}} \approx 2600$ km/s farther out.

\begin{table*}
 \centering
\caption{Prominent Emission Lines Properties from the SDSS Spectrum}
{\tiny
\begin{tabular}{cccccccc}
 \hline

Line &   Broad Component &  Full Line & Line Shift & Redward Asymmetry & Absorption  &  Absorption Velocity  &  Notes\\
       & FWHM (km/s)   &  FWHM (km/s)& km/s & $A$  &  FWHM (km/s) & km/s  &\\
\hline
 Mg II  & 3060 &  4110 & 440 & 0.32 &  305/370    & -2632/-2609   &  1 \\
 C III  & 3702 &  4290 & 430 & 0  &  0    & ...   &  .. \\
 C IV  & 3070 &  3870 & 265 & 0.15/0.24  &  1200    & -2663   &  2 \\
\hline
\end{tabular}}
\par
\footnotesize{1. The two absorption widths are from resolved doublet
2. The larger asymmetry is from \cite{and87} as seen in Figure 3.}
\end{table*}
\par The UV emission lines were decomposed into a broad, a very broad
and narrow components as has been previously documented in the
analysis of eigenvector 1 and the Population A/B dichotomy used to
classify quasar spectrum \citep{sul07}. It is demonstrated that the
line profiles are typical of RLQ, Pop. B sources which are at the
opposite end of eigenvector 1 from the BALQSOs (Pop. A sources).
From Table 1, the peaks of the CIV, CIII] and MgII broad components
show small redshifts relative to the quasar systemic velocity
consistent with Pop B sources \citep{sul07}. Secondly, here is a
tendency for redward asymmetric profiles which is also associated
with RLQ, Pop. B sources. Using the measure for asymmetry from
\cite{wil95}, $A$, that is defined in terms of the full width half
maximum, FWHM, in $\AA$; the midpoint of an imaginary line
connecting a point defined at 1/4 of the peak flux density of the
BEL on the red side of the BEL to 1/4 of the peak flux density on
the blue side of the BEL, $\lambda_{25}$, and a similar midpoint
defined at 8/10 of the flux density maximum, $\lambda_{80}$, as
\begin{equation}
A = \frac{\lambda_{25} - \lambda_{80}}{FWHM} \;.
\end{equation}
A positive value of $A$ means that there is excess flux in the red
broad wing of the emission line. The redward asymmetry, especially
of CIV, has been previously associated with radio loud quasars
\citep{pun10}. The values in Table 1 are large even for RLQs.
Finally, the small intensity ratio Al III 1860 / CIII] 1909
$\approx$ 0.08 is typical of Pop. B sources \citep{bac04}. The
designation as a Pop. B source is consistent with the Mg II and CIV
absorption widths in Table 1. The narrow widths are typical of
narrow absorption lines not broad absorption lines.

\par Using the MgII broad component width from Table 1, we obtain
mass estimates for the central black hole $\log{M_{bh}}=8.89$ and
$\log{M_{bh}}=8.99$ from the methods of \cite{she12} and
\cite{tra12}, respectively. Combining these estimates with $L_{bol}$
from Section 1 indicates an Eddington rate, $L_{bol} =0.48\pm0.10
L_{\mathrm{Edd}}$. This is very high for a Pop. B source and is more
typical of a Pop. A, radio quiet source. Thus, 3C 270.1 is extreme
in all its properties, a very strong jet and a very luminous
accretion disk.
\section{Conclusion}
In the first section, the EUV spectrum of 3C 270.1 was found to have
$\alpha_{EUV} =2.98\pm 0.15$. It was shown that 3C 270.1 has all the
properties associated with the one of the most powerful RLQ jets in
the known Universe. An argument was made that the
$\log[\overline{Q}]$ - $\alpha_{EUV}$ scatter plane (Figure 2) can
be used to estimate a confidence interval for real time jet power,
$Q(t)$. Unlike other correlations in the literature involving
$\overline{Q}$, such as with emission line properties as in
\cite{wil99}, it was noted that both the EUV and $Q(t)$ originate
from near the central black hole and the correlation in the
$\log[\overline{Q}/L_{bol}]$ - $\alpha_{EUV}$ scatter plane
therefore indicates a nearly simultaneous causal contact between the
two dynamical processes. If not for this circumstance, the estimator
for $Q(t)$ would be unjustified. The method developed here might
have application to other extreme quasars in future studies. In the
second section, it was shown that the CIV and OVI absorption had $BI
=0$. Thus, 3C 270.1 is not a broad absorption line quasar as
classically defined, but displays the well known associated
absorption that occurs in many RLQs.
\par The details of the correlation found in Figure 2 was elucidated
in two previous studies. Firstly, a correlation induced by larger
black hole masses and lower accretion rates in RLQs was ruled out
empirically as a plausible explanation due to the indistinguishable
SED peak in RLQs and RQQs \citep{pun14}. A partial correlation
analysis in \citet{pun15} indicates that the fundamental physical
correlation amongst quantities is between $Q/L_{\mathrm{bol}}$ and
$\alpha_{EUV}$. This correlation is not explained by the
\cite{lao14} disk wind. However, as discussed in \citet{pun14} the
most likely explanation of the correlation, jets from magnetic flux
in the inner accretion disk, can coexist with these winds.
Consistent 3-D MHD numerical models are those in which an annular
region ($\sim$ 2-4 black hole radii wide) of the innermost accretion
flow, adjacent to the black hole, is perforated by islands of large
scale poloidal (vertical) magnetic flux. The rotating flux
distribution associated with the magnetic islands is the source of
the relativistic jet and it also displaces the EUV emitting gas,
thereby causing an EUV deficit that is correlated with $Q(t)$
\citep{pun15}. The degree of magnetization of the inner disk
determines scaling relations for jet power and the EUV decrement. In
particular, based on Figure 4 of \cite{pun15}, the EUV deficit and
jet power of 3C 270.1 indicates that $\approx 50\%$ of the innermost
accretion flow is displaced by the magnetic islands.


\begin{thebibliography}{99}
\bibitem[\protect\citeauthoryear{Akujor et al.}{1994}]{aku94} Akujor, C., et al.. 1994, \emph{Astron.\& Astrophys. Suppl.}\textbf{105}
247
\bibitem[\protect\citeauthoryear{Aldcroft et al.}{1994}]{ald94}Aldrcoft, T., Bechtold, J., Elvis, M. 1994 \emph{Astrophys. J. Suppl.} \textbf{95}
1
\bibitem[\protect\citeauthoryear{Anderson et al}{1987}]{and87}Anderson, S., Weymann, R., Foltz, C., Chaffee, F. 1970 \emph{Astron. J.} \textbf{94} 278
\bibitem[\protect\citeauthoryear{Bachev et al.}{2004}]{bac04} Bachev, R., Marziani,
P., Sulentic, J.~W., et al.\ 2004, \emph{Astrophys. J.} \textbf{617}
171
\bibitem[\protect\citeauthoryear{Becker et al.}{2001}]{bec01} Becker, R., et al. 2001 \emph{Astrophys. J. Supp.} \textbf{135} 227
\bibitem[\protect\citeauthoryear{Blundell and Rawlings}{2000}]{blu00} Blundell, K., Rawlings, S. 2000 \emph{Astron. J.} \textbf{119} 1111
\bibitem[\protect\citeauthoryear{Cardelli et al.}{1989}]{car89} Cardelli, J., Clayton, G., Mathis, J. 1989 \emph{Astrophys. J.} \textbf{345}
245
\bibitem[\protect\citeauthoryear{Davis and Laor}{2011}]{dav11} Davis, S., Laor, A. 2011, \emph{Astrophys. J.} \textbf{728} 98
245
\bibitem[\protect\citeauthoryear{deVries et al.}{2006}]{dev06} deVries, W., Becker, R., White, R. 2006, \emph{Astron. J.} \textbf{131} 666
\bibitem[\protect\citeauthoryear{Garrington et al.}{1991}]{gar91} Garrington, S., Conway, R.,  Leahy, J.,
1991 \emph{Mon. Not. R. Astr. Soc.} \textbf{250} 173
\bibitem[\protect\citeauthoryear{Gibson et al.}{2009}]{gib09} Gibson, R. et al 2009 \emph{Atrophys. J.} \textbf{692} 758
\bibitem[\protect\citeauthoryear{Gregg et al.}{2006}]{gre06} Gregg, M., Becker, R., de Vries, W., 2006 \emph{Astrophys. J.} \textbf{641} 210
\bibitem[\protect\citeauthoryear{Jackson and Rawlings}{1997}]{jac97}Jackson, N., Rawlings, S. 1997 \emph{Mon. Not. R. Astr. Soc.} \textbf{286}
241
\bibitem[\protect\citeauthoryear{Knigge et al.}{2008}]{kni08}Knigge, C., Scaringi, S., Goad, M., Cottis, C. 2008 \emph{Mon. Not. R. Astr. Soc.} \textbf{386}
1426
\bibitem[\protect\citeauthoryear{Laor and Davis}{2014}]{lao14}Laor, A., Davis, S. 2014 \emph{Astrophys. J.} \textbf{428} 3024
\bibitem[\protect\citeauthoryear{Lonsdale et al.}{1993}]{lon93}Lonsdale, C., Barthel, P., Miley, G. 1993 \emph{Astrophys. J. Supp.} \textbf{87} 63
\bibitem[\protect\citeauthoryear{Punsly}{2005}]{pun05}Punsly, B. 2005 \emph{Astrophys. J. Lett.} \textbf{623} 9
\bibitem[\protect\citeauthoryear{Punsly}{2006}]{pun06}Punsly, B. 2006 \emph{Astrophys. J.} \textbf{647} 886
\bibitem[\protect\citeauthoryear{Punsly}{2007}]{pun07}Punsly, B. 2007 \emph{Mon. Not. R. Astr. Soc. Lett.} \textbf{374} 10
\bibitem[\protect\citeauthoryear{Punsly}{2010}]{pun10}Punsly, B. 2010 \emph{Astrophys. J. Lett.} \textbf{713}
232
\bibitem[\protect\citeauthoryear{Punsly}{2014}]{pun14}Punsly, B. 2014 \emph{Astrophys. J. Lett.} \textbf{797} 33
\bibitem[\protect\citeauthoryear{Punsly}{2015}]{pun15}Punsly, B. 2015 \emph{Astrophys. J.} \textbf{806}
47
\bibitem[\protect\citeauthoryear{Reid et al.}{1995}]{rei95}Reid, A., et al. 1995 \emph{Astron.\& Astrophys. Suppl.}\textbf{110} 213
\bibitem[\protect\citeauthoryear{Richards et al.}{2011}]{ric11}Richards, G. et al.2011 \emph{Astron. J.} \textbf{141} 167
\bibitem[\protect\citeauthoryear{Shankar et al.}{2008}]{sha08}Shankar, F., Dai, X., Sivakoff, G., 2008\emph{Astrophys. J.} \textbf{687} 859
\bibitem[\protect\citeauthoryear{Shen and Liu}{2012}]{she12} Shen, Y., \& Liu, X.\ 2012,
\emph{Astrophys.J.} \textbf{753} 125
\bibitem[\protect\citeauthoryear{Shull et al.}{2012}]{shu12}Shull, M., Stevans, M., Danforth, C. 2012\emph{Astrophys. J.} \textbf{752}
162
\bibitem[\protect\citeauthoryear{Sulentic et al.}{2007}]{sul07}Sulentic, J.~W.,Bachev, R., Marziani, P., Negrete, C.~A., \& Dultzin, D.\ 2007, \emph{Astrophys. J.},
\textbf{666}, 757
\bibitem[\protect\citeauthoryear{Telfer et al.}{2002}]{tel02} Telfer, R., Zheng, W., Kriss, G., Davidsen, A. 2002 \emph{Astrophys. J.} \textbf{565} 773
\bibitem[\protect\citeauthoryear{Trakhtenbrot
\& Netzer}{2012}]{tra12} Trakhtenbrot, B. and Netzer, H.\ 2012,
\emph{Mon. Not. R. Astr. Soc.}  \textbf{427}, 3081
\bibitem[\protect\citeauthoryear{Weymann et al.}{1991}]{wey91} Weymann, R.J., Morris, S.L., Foltz, C.B., Hewett, P.C. 1991, \emph{Astrophys. J.}  \textbf{373}, 23
\bibitem[\protect\citeauthoryear{Willott et al.}{1999}]{wil99}Willott, C., Rawlings, S., Blundell, K., Lacy, M. 1999 \emph{Mon. Not. R. Astr. Soc.} \textbf{309} 1017
\bibitem[\protect\citeauthoryear{Wills et al.}{1995}]{wil95} Wills, B. et al 1995, \emph{Astrophys. J.} \textbf{437} 139
\bibitem[\protect\citeauthoryear{Zheng et al.}{1997}]{zhe97} Zheng, W. et al. 1997 \emph{Astrophys. J.} \textbf{475} 469
\end{thebibliography}
\end{document}